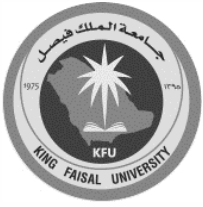

**المجلة العلمية لجامعة الملك فيصل**
The Scientific Journal of King Faisal University

العلوم الأساسية والتطبيقية
Basic and Applied Sciences

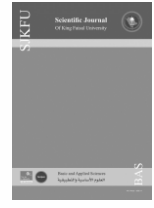

# An Overview of Query Processing on Crowdsourced Databases

Marwa B. Swidan[1], Ali A. Alwan[1], Yonis Gulzar[2], Abedallah Zaid Abualkishik[3]

[1] International Islamic University Malaysia, Kuala Lumpur, Malaysia
[2] College of Business Administration, King Faisal University, Al Ahsa, Saudi Arabia
[3] American University in The Emirates, Dubai, United Arab Emirates

## ABSTRACT

Crowd-sourcing is a powerful solution for finding correct answers to expensive and unanswered queries in databases, including those with uncertain and incomplete data. Attempts to use crowd-sourcing to exploit human abilities to process these expensive queries using human workers have helped to provide accurate results by utilising the available data in the crowd. Crowd-sourcing database systems (CSDBs) combine the knowledge of the crowd with a relational database by using some variant of a relational database with minor changes. This paper surveys the leading studies conducted in the area of query processing with regard to both traditional and preference queries in CSDBs. The focus of this work is on highlighting the strengths and the weakness of each approach. A detailed discussion of current and future trends research associated with query processing in the area of CSDBs is also presented.

## 1. Introduction

In many contemporary database applications, queries cannot always be optimally answered through traditional techniques. There are many causes for this, including imprecise or uncertain information. Examples of such applications include translation, handwriting recognition, image understanding, and web databases. Crowd-sourced platforms have become an effective solution for finding answers to various types of queries by exploiting the knowledge, ideas, experiences, and skills of crowd workers to process information and obtain accurate responses. A hybrid human/machine system is defined as a process that integrates crowd workers with computer systems to generate high-quality results (Parameswaran *et al.*, 2011; Li *et al.*, 2016; Swidan *et al.*, 2018; Bhaskar *et al.*, 2020; Lian *et al.*, 2020; Swidan *et al.*, 2020). A crowd-sourced system consists of three main components, which work together to provide users with the best answer to a particular query: (i) a requester, who submits a query to the crowd and waits for the answer to be generated; (ii) a crowd platform, which contains the query execution engine that is responsible for managing the user's query and retrieving answers; and (iii) workers, who are responsible for working on the queries and delivering the results to the requester via the crowd platform (Difallah *et al.*, 2015; Chai *et al.*, 2019).

Figure 1 shows a detailed illustration of the structure of a typical crowd-sourced platform. The requester submits the query to the crowd, specifying the requirements for accomplishing the task. The crowd divides the given query (task) into many sub-tasks, also called micro-tasks or human intelligent tasks (HITs). The platform then assigns these HITs to workers, who each agree to carry out a task. Assigning tasks to workers is the most critical aspect of query processing in the crowd-sourced system, since workers attempt to generate a result based on the user's query, and the quality of the result is strongly influenced by the quality of the selected workers. Not all workers are suitable to accomplish a given task, and the decision to select the most appropriate worker is therefore made by the requester (Franklin *et al.*, 2011; Difallah *et al.*, 2015; Chai *et al.*, 2019; Lian *et al.*, 2020).

**Figure 1: Query processing in the crowd-sourcing system**

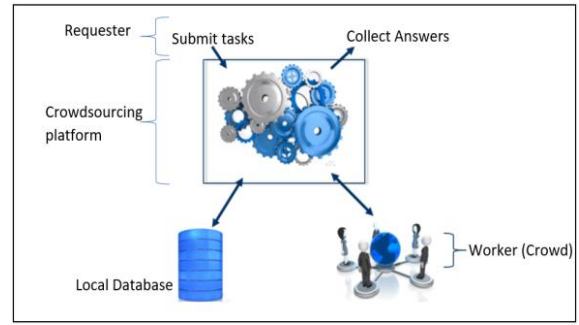

A CSDB contains a massive amount of heterogeneous data, which are stored in various online locations accessible by the crowd. These data may be in many different forms, such as web data, text, images, audio, and personal data (Li *et al.*, 2016; Lian *et al.*, 2020). A CSDB uses many of the functions of traditional database systems and SQL operators to process complex queries, and extends some of the traditional operators and adds new crowd operators to a database management system (DBMS) (Parameswaran *et al.*, 2011; Li *et al.*, 2016). Several crowd-sourced systems have been developed with the aim of extending a traditional database system to work on crowd platforms. These extensions support a wider range of types of query that rely on the power of human knowledge, and include CrowdDB (Franklin *et al.*, 2011), Qurk (Marcus *et al.*, 2011) and Deco (Parameswaran *et al.*, 2012a). These crowd-sourced systems are associated with specific crowd-sourced marketplaces, for instance Amazon Mechanical Turk (Franklin *et al.*, 2011) and CrowdFlower (Li *et al.*, 2016). There are many differences between traditional and crowd-sourced databases. Firstly, traditional databases can be considered as a closed world in terms of query processing, meaning that data that are not in the database are treated as false or non-existent. Secondly, traditional databases are extremely literal; for example, the DBMS assumes that information entered by a user has been appropriately cleaned and validated before being entered into the database. Query processing via crowd databases is typically influenced by three critical factors: monetary cost, time latency, and the accuracy of the results (Parameswaran *et al.*, 2011; El Maarry *et al.*, 2015; Lee *et al.*, 2016; Li *et al.*, 2016; Chai *et al.*, 2019; Bhaskar *et al.*, 2020).

### 1.1. Monetary Cost:

Crowd services are not free, and every task presented by a requester incurs some monetary cost. In other words, when the requester collects an accepted answer, he or she needs to pay the worker. The requester therefore needs to know in advance the estimated monetary cost for each task before assigning it to a worker. Answering user queries by exploiting a crowd platform can be expensive when the number of tasks is large (Lee *et al.*, 2016).

### 1.2. Time Latency:

This factor denotes the period between the submission of the task by the requester and the collection of the answers from the crowd. When submitting a task, the requester can add some constraints to be considered by the workers when processing the task, including a time limit for completing the task, the maximum time the crowd can see the task on the platform (i.e. an expiration time for the task), and the estimated monetary cost that the requester is willing to pay for the given task (Li *et al.*, 2016). In certain cases, the workers may be unreliable, or may attempt to hold the task for a long time without giving valid answers. These issues can significantly increase the time latency of processing the tasks.

### 1.3. Result Accuracy:

Sometimes, the results produced by the crowd workers are not accurate, since humans often tend to make errors. There are various causes for these errors: some may be intentional and some unintentional. Unintentional errors can occur if the worker misunderstands the assigned task, while deliberate errors are made by malicious workers, who may submit random answers just to collect payment. In both cases, the accuracy of the result is compromised.

It can be concluded that in a crowd environment, there is a trade-off between the accuracy of the result and the monetary cost of accomplishing the task. In certain scenarios, the requester is more concerned about the monetary cost, which in turn leads to a reduction in the expected quality of the result. Conversely, the requester may be more concerned with the quality of the answer, which may result in a higher monetary cost. The most important aspect is that workers have different levels of expertise, skills and training, which may make some of them unable to achieve a high-quality result, and workers are sometimes incapable of working on certain tasks (Li *et al.*, 2016). In this paper, we review some leading studies of both traditional and preference queries in the area of CSDBs. We also highlight and explain the differences between the processing of queries in both traditional databases and CSDBs.

The remainder of this paper is organised as follows. Section 2 reviews the literature on traditional queries in CSDBs. Section 3 discusses studies of preference queries in a CSDB, including some leading studies related to preference queries in traditional databases. Section 4 contains a detailed discussion and an outline of future research challenges. Finally, the conclusion to the paper is presented in Section 5.

## 2. Traditional Queries in Crowd-Sourced Databases

Traditional queries generate results by strictly relying on the conditions given in the query, and a nil result may be returned if no tuple satisfies the given conditions. In the same way as traditional databases, CSDBs use data manipulation functions or traditional queries (such as select, aggregate, maximum, and average), except that these rely on crowd workers. This section discusses the approaches used for traditional queries in a crowd-sourced database.

### 2.1. Selection Operators:

The selection operator is used to retrieve any data from the database that satisfy the conditions given in the user query. The filter in a CSDB causes the human workers to identify a set of tuples that satisfy the query constraints, similarly to a 'SELECT' operator in a traditional database. For example, a user may ask the crowd to select historical movies from a movie database. The CrowdDB (Franklin *et al.*, 2011) and Qurk (Marcus *et al.*, 2011) systems have introduced various types of heuristic filtering strategies (such as majority voting over a fixed number of workers), but have not incorporated ways to implement optimal filtering strategies, these systems attended by optimising the monetary cost. The authors of CrowdDB (Franklin *et al.*, 2011) optimised the join, compare and fill operators, while the authors of Qurk (Marcus *et al.*, 2011) studied the join and sort operators. Deco (Parameswaran *et al.*, 2012a) focuses on missing value records in the database. In contrast, the CrowdScreen (Parameswaran *et al.*, 2012b) system uses only the select (filter) operator, and aims to determine a strategy that minimises the number of tasks submitted to the crowd by taking into consideration the total error, which must be less than a given threshold. CrowdFind (Sarma *et al.*, 2014) attempts to estimate the cost and latency of filtering via the crowd. The system allows for a batch of questions to be asked in parallel. CrowdOP (Fan *et al.*, 2015) aims to optimise more general operators than CrowdFind, and attempts to minimise the cost and latency of the select, join, and complex operators.

### 2.2. Sorting Operators:

The sort operator uses one or more attributes from a relational database as a basis for ordering all the tuples in the database. CSDBs exploit human computation to apply the sort operator. For instance, several photos of a restaurant may be provided and the crowd may be asked to sort these based on which photo illustrates the restaurant best; the better photos may then be shown on its website. To apply a sorting operator in a crowd-sourced system, strategies are developed to change the method of comparison used in a traditional sorting algorithm to a new approach that depends on workers to evaluate the comparison and to decide which value is better. The CrowdDB system implemented the CrowdCompare operator in the form of a CROWDEQUAL function, which takes two values and asks the crowd to identify which pairs are equal. The CROWDORDER function is used to rank or order the results (Franklin *et al.*, 2011). In Qurk (Marcus *et al.*, 2011), the rankProducts function is implemented to perform a comparison between many tuples in order to sort them.

### 2.3. Filling Operators:

Most crowd-sourced systems attempt to support queries in an incomplete database. Many systems have implemented operators that generate the values missing from the database to identify the best results. CrowdDB (Franklin *et al.*, 2011) implements the CrowdProbe operator, which aims to obtain missing information from workers for crowd-sourced tables. CDB (Li *et al.*, 2017) introduced two built-in keywords for data gathering, FILL and COLLECT. FILL asks the crowd to find and fill in the values of attributes with rational values, for example filling in missing values of the *manufacturer* attribute for cars, whereas COLLECT asks the crowd to list the 10 best cars.

### 2.4. Join Operators:

The join operator combines at least two tuples from the same or different databases to obtain results based on certain conditions. A wide range of approaches for performing a join operation in a CSBD have been presented in previous studies. In CrowdDB (Franklin *et al.*, 2011), the CrowdJoin operator creates a join based on an indexed nested loop between two tables, at least one of which must be a crowd-sourced table. In Qurk (Marcus *et al.*, 2011), the crowd is entrusted with determining whether tuples from two relations are matched in terms of at least one attribute. A naive implementation of this join would ask the workers about every possible pairing of tuples between the relations. In CrowdOP (Fan *et al.*, 2015) the Crowd-Powered Join (CJOIN) operator was introduced to leverage the crowd in order to merge objects from two sources, as indicated by the specific constraints. For join queries, the CrowdOP incorporates the FILL and JOIN operators into a single operator. This algorithm asks each crowd worker to fill in some incomplete values, and then combines these values using a join operator. For instance, to match the images of mobile phones with their reviews, crowd workers are first asked to fill in the brand of each mobile phone and then to match mobile phones only with the same brand. Wang *et al.* (2013) focused on a crowd-sourced join query for entity resolution, which aimed to find all pairs of matching tuples between two collections of tuples by asking the crowd to decide which pairs of tuples were matched. CDB (Li *et al.*, 2017) introduced two algorithms called Crowdequal and Crowdjoin to carry out join queries in a CSDB. Table 1 summarises the CSDBs examined in this section.

Table 1: Summary of crowd-sourced database Systems

| Author/ Year | CSDB system | Query Operators | Performance Metrics |
|---|---|---|---|
| Franklin *et al.*, 2011 | CrowdDB | Join, compare, fill | Cost |
| Marcus *et al.*, 2011 | Qurk | Join, sort, select | Cost |
| Parameswaran *et al.*, 2012a | Deco | Fill | Cost |
| Parameswaran *et al.*, 2012 b | CrowdScreen | Select | Cost, latency |
| Wang *et al.*, 2013 | Framework | Join | Cost, latency, quality |
| Sarma *et al.*, 2014 | CrowdFind | Select | Cost, latency |
| Fan *et al.*, 2015 | CrowdOP | Select, join, fill | Cost, latency |
| Li *et al.*, 2017 | CDB | Select, join, fill, collect | Cost, latency, quality |

## 3. Preferences Queries In Crowd-Sourcing Databases

A preference query is another essential type of query that has been incorporated into a large number of modern database applications. Preference queries aim to relax the query conditions by identifying results that rely on the user's preferences for one tuple over others. The growing complexity of contemporary database applications and the need to support users with different preferences have further increased the need for new types of query operators to be combined into these systems. Many applications in various domains have significantly benefited from these types of queries, for example web services, multi-criteria decision-making applications, digital libraries, multimedia retrieval, and recommender systems. Due to the importance of preference queries in many database applications, many methods for evaluating preferences have been proposed. These techniques rely on the skyline and top-*k* techniques.

### 3.1. Top-*k* Queries in Crowd-Sourced Systems:

The top-*k* approach ranks the tuples of a dataset by collecting the values of each attribute of a tuple into a single value using a monotonic ranking function *F*. The best *k* tuples are ranked based on the best scoring value using a function *F* (Chaudhuri *et al.*, 1999; Chang *et al.*, 2002). Top-*k* queries have several benefits: (i) they are transitive, meaning that when one tuple *p* dominates another, *q*, and *q* dominates another tuple *r*, then *p* dominates *r*; (ii) the output of top-*k* queries is controlled by simply varying the value of *k*, since this represents the number of tuples to be retrieved for the user; (iii) no exhaustive pairwise comparison between the individual attributes of tuples is necessary when identifying the results of top-*k* queries. Top-*k* queries are subject to a monotonic ranking function defined by users, meaning that different top-*k* results can be generated based on different preference functions. Moreover, when determining the most interesting tuple of a query, it is not easy to analyse the data, since different scores and functions may give different results (Gumaei *et al.*, 2017; Kontaki *et al.*, 2010). The results of top-*k* queries are negatively influenced by incomplete data, meaning that if the tuples have missing values in one or more dimensions, the answers to a query may not be the same as for complete data. Furthermore, relying on the application scenarios, in the case where the weights of the linear criterion are known in advance, for future queries might be pre-ranked the results. Otherwise, it may not be feasible to pre-compute all possible combinations, since the weights are dynamically allocated (Kontaki *et al.*, 2010; Gumaei *et al.*, 2017; Chang *et al.*, 2002). To perform a top-*k* query in a CSDB, the system sorts the whole database and then returns the top value(s). Crowd-sourced top-*k* algorithms ask the crowd to compare tuples and to detect the top-*k* tuples based on the results of this crowd-sourced comparison. However, these algorithms can lead to high costs, as unnecessary comparisons are performed during the process of the top-*k* query. The task is processed in two ways: the first is a single choice, in which each task involves comparing two tuples, and the second is a rating, in which multiple tuples are selected and the crowd is asked to assign a rate for each tuple. Many methods relying on tournament sorting have been presented in the literature to implement crowd-based top-*k* algorithms.

Polychronopoulos *et al.* (2013) presented a solution based on a human-powered top-k method in which the crowd is asked to rank tuples directly and to aggregate them using a median rank aggregation algorithm. This leads to the final top-*k* result being identified based on the judgments of human workers. This approach allows workers to examine several items at a time, with no prior knowledge about the errors made by the human workers. It adapts dynamically to the varying difficulty of comparing items and the existence of spammers. However, top-*k* lists that are larger than the size of the ranking tasks given to the crowd cannot be tackled. The method presented by Lin *et al.* (2017) concentrates on reducing the uncertainty of a top-*k* ranking using a crowdsourcing approach. The issue of uncertainty leads to low-quality results, and algorithms that try to identify the best tuple pairs via crowdsourcing leading to highest quality improvement with the crowd-sourced task of a limited budget have been proposed. Ciceri *et al.* (2015) aimed to reduce the expected residual uncertainty of the results by using a set of questions that, within an allowed budget for crowdsourcing, could lead to a unique ordering of the top-*k* results. De Alfaro *et al.* (2016) proposed a method for top-*k* queries that leverages the knowledge obtained from crowd comparisons to reduce the set of candidate tuples of the top-*k* list. The idea in this approach involved controlling the size of the crowd comparison tasks, decoupled from the size of the top-*k* results list.

Lee *et al.* (2017) studied the issue of top-*k* queries in a CSDB, and developed a method called crowd-enabled top-*k* queries. A novel framework was proposed that aimed to increase the monetary cost when the latency is constrained. This strategy consisted of a two-phase parameterised framework with two parameters, called buckets and ranges. Three methods, called greedy, equi-sized, and dynamic programming, were used to identify the buckets and ranges. By combining these three methods in each phase, four algorithms were designed, called GdyBucket, EquiBucket, EquiRange, and CrowdK. Lastly, Kou *et al.* (2017) introduced a new framework for processing top-*k* queries in crowdsourced systems with the aim of minimising the monetary cost and latency, and maximising the quality of the results. Their framework attempted to reduce the monetary cost by decreasing the number of comparisons, and to improve the accuracy by introducing a judgment model that allowed for pairwise preference judgments. Table 2 summarises the existing approaches to top-*k* queries discussed in this section.

**Table 2: Summary of previous approaches of top-k queries in a crowd-sourcing database**

| Author/ Year | Approach | No. of Workers | K values | Database Type | Performance Metrics |
|---|---|---|---|---|---|
| Polychronopoulos et. al., 2013 | Algorithm | 3-15 | 1-5 | Synthetic, real | Cost, latency |
| Lin et. al., 2017 | A novel pairwise crowdsourcing model | - | 10 – 30 | Real | Quality |
| Ciceri, et. al., 2015 | Heuristic algorithm | Multiple workers | 1- 10 | Synthetic, real | Latency |
| De Alfaro et. al., 2016 | Crowd-top-*k* | 3 | 5-50 | Synthetic, real | Cost, latency, quality |
| Lee et. al., 2017 | CrowdK | 3, 5, 7, 9 | 2 – 20 | Synthetic, real | Cost, latency, quality |
| Kou et. al., 2017 | SPR | - | 1, 5, 10, 15, 20 | Real | Cost, latency, quality |

### 3.2. Skyline Queries in Crowd-sourced Databases:

The skyline technique retrieves the skyline S in such a way that any tuple in S is not dominated by any other tuples in the database (Gulzar *et al.*, 2017). Skyline queries have many powerful characteristics, for example: (i) each attribute is evaluated independently through the process of pairwise comparison, and a user-defined ranking function is not required; (ii) the skyline results are identified based on the real values of the data in the database; (iii) the scale of attributes does not affect the skylines, as the comparison process relies solely on the present value of each attribute; (iv) the integration of the skyline operator into the SQL query processor is very simple and straightforward; (v) this approach intuitively performs multi-objective query operations; (vi) it has the property of transitivity, which is the main theme of the skyline technique; and (vii) it can be used as an effective operator for aggregate operations (Kontaki *et al.*, 2010; Kou *et al.*, 2017). However, skyline queries have some weaknesses, for instance: (i) there is no control on the size of the skylines, and in the worst case, all the tuples in the database may be retrieved as skylines; (ii) the computational complexity of the skyline process is strongly influenced by the number of attributes and the size of the database; and (iii) due to the enormous number of skylines, the user needs to examine numerous skylines to determine the most suitable tuples to be selected.

In the era of the Internet of Things, where enormous amounts of data are collected from different sensors in remote locations and are transmitted between devices, there is a high probability that certain data items have missing values in one or more attributes, meaning that the database is incomplete. Processing skyline queries in an incomplete database can raise some serious challenges; for example, the *transitivity property* of skyline technique is unlikely to hold for a database with incomplete data. Loss of the *transitivity property* may lead to the domination test becoming *cyclic* due to the fact that some tuples are incomparable with each other, meaning that the skyline result could be nil (Khalefa *et al.*, 2008; Gulzar *et al.*, 2016a; Gulzar *et al.*, 2018b).

Many algorithms have been suggested for processing skyline queries in incomplete databases, such as Bucket and Iskyline (Khalefa *et al.*, 2008), Sort-based Incomplete Data Skyline (SIDS) (Bharuka *et al.*, 2013), Incoskyline (Alwan *et al.*, 2016), Select-Partition-Rank (Kou *et al.*, 2017) and SOBA (Lee *et al.*, 2016). Bucket and Iskyline (Khalefa *et al.*, 2008) were among the first attempts at processing skyline queries in a database with incomplete data. Iskyline exploits the clustering technique to process skylines, whereas the SIDS algorithm (Bharuka *et al.*, 2013) and Select-Partition-Rank framework (Kou *et al.* 2017) use a sorting technique to rearrange the tuples such that tuples with a high probability of being part of the skyline are kept at the top of the sorted list. In the Incoskyline (Alwan *et al.*, 2016) algorithm, a virtual tuple called *k*-dom is created from the local skylines of each group created after clustering. Incoskyline aims to reduce the execution time and to decrease the domination tests required during the skyline process. Lastly, the SOBA technique (Lee *et al.*, 2016) combines the sorting and clustering approaches to process skyline queries in incomplete databases.

The issue of processing skyline queries in an incomplete database has been further investigated to include more complicated databases, such as distributed and cloud architectures. Gulzar *et al.* (2019a) processed skyline queries over cloud-based incomplete databases, where data were divided horizontally and stored in different data centres at different locations. Moreover, Alwan *et al.* (2017) proposed an algorithm for processing skyline queries on incomplete distributed databases. Their algorithm, named *Jincoskyline*, uses a vertically divided database stored in different relations. However, little attention has been paid to the issue of estimating the missing values of skylines. The only work that has addressed this issue has been the study by Alwan *et al.* (2018), which aimed to capture the missing values before identifying the final skylines. Most of the skyline techniques mentioned above were designed to work on traditional databases, and the topic of processing skyline queries for a crowd-sourced enabled incomplete database has not been investigated by many researchers in the database community. Several unique features distinguish CSDBs from traditional databases, and it may be unwise to apply approaches designed for traditional databases directly to crowd-sourced databases. Some previous studies have focused on designing approaches for the prediction of missing values from skylines, with the aim of finding a balance between the monetary cost, the time latency and the accuracy of the results.

In the following, we discuss some leading studies that focus on the processing of skyline queries on crowd-sourced enabled incomplete databases. This discussion elaborates the strengths and weaknesses of these works.

Asudeh *et al.* (2015) highlighted the issue of identifying Pareto-optimal tuples in CSDBs. Their approach is useful when the Pareto-optimal tuples do not have clear attributes and preference relations are strict partial orders. The authors introduced an iterative question selection algorithm that is instantiated in different methods based on the ideas of candidate questions, macro-ordering, and micro-ordering. Lofi *et al.* (2013) and El Maarry *et al.* (2015) suggested a hybrid method that combined dynamic crowd-sourcing with heuristic techniques. The aim was to use a set of heuristic techniques to estimate the incomplete values for all tuples in the database. Workers are used in some cases, to improve the quality of the estimated values. A new approach called CrowdSky was proposed by Lee *et al.* (2016), which processed skyline queries for a crowd-sourced enabled incomplete database. The focus was on decreasing the monetary cost, minimising the latency, and improving the quality of the results. Lee *et al.* (2016) assumed that although an initial database may be complete, the data in the database might be insufficient to provide an accurate answer for skyline queries. Hence, there may be a need to consult the crowd to estimate this missing information from a set of virtual attributes to be created when answering the query. To the best of our knowledge, the study by Miao *et al.* (2019) is the most recent to highlight the issue of skyline queries on crowd-sourced enabled incomplete databases. These authors proposed a framework for solving the skyline problem with a partially complete database using crowd-sourcing, called *Bayescrowd*. Both *BayesCrowd* and *CrowdSky* used only the crowd to estimate the missing values. However, the researchers assumed that all missing values could be estimated by utilising CSDBs. Table 3 summarises the existing skyline techniques for the CSDBs discussed in this section.

Table 3: Summary of previous approaches of skyline techniques in crowd-sourcing database

| Author /Year | Approach | Data Distribution | Database Type | No. of Dimensions | Missing Data | Performance Metrics |
|---|---|---|---|---|---|---|
| Lofi *et al.*, 2013 | Heuristic *KNN* | Independent, anti-correlated | Real | 6-8 | 1% -20% | Latency, cost, quality |
| EL Maarry *et al.*, 2015 | Heuristic min value | Independent, anti-correlated | Real | 6-8 | 1% -20% | Cost, quality |
| Asudeh *et al.*, 2015 | Macro-ordering and micro-ordering heuristics | - | Real, synthetic | 10 | 0% | Latency, cost, quality |
| Lee *et al.*, 2016 | CrowdSky | Independent, anti-correlated | Real, synthetic | 2-5 | 1 – 3 dimensions | Latency, cost, quality |
| Miao *et al.*, 2019 | Bayescrowd | - | Real, synthetic | 9-11 | 20%-50% | Latency, cost, quality |

## 4. Discussion and Future Research Challenges

From the literature, we can conclude that the monetary cost, time latency, and accuracy of results are the factors with the most impact on query processing in CSDBs. Most previous studies have concentrated on the design of query processing operators for CSDBs, with the aim of improving the accuracy of the results for a reasonable cost and time. The cost and accuracy of results are among the most critical factors considered in most previous works (Franklin *et al.*, 2011; Parameswaran *et al.*, 2012b; Lofi *et al.*, 2013; Wang *et al.*, 2013; Swidan *et al.*, 2018; Chai *et al.*, 2019; Lian *et al.*, 2020; Swidan *et al.*, 2020). Numerous optimisation methods have been applied to reduce the monetary cost of extracting and generating data from the crowd, such as decreasing the number of questions submitted to the crowd (Lee *et al.*, 2016) or setting a specific budget for processing queries (Ciceri, *et al.*, 2015). In contrast, reducing the number of questions and limiting the monetary cost is relatively easy for traditional queries. This is because most traditional query operators access the underlying data in the database, and only certain attributes can be accessed. However, it should be noted that it is very challenging to reduce the monetary cost for preference queries such as top-*k* and skyline, since these types of queries need to scan the entire database, and a complex process is required to derive the query result. For a top-*k* query, a monotone function needs to be formed that aggregates the values of all dimensions and sorts the result in a particular order. Similarly, for skyline queries, exhaustive pairwise comparisons need to be performed to determine the skylines of the database. This process requires accessing the entire database, meaning that the search space is extremely large. Many studies have tried to focus on developing query algorithms that reduce the cost and time latency of task, but less attention has been paid to the quality of the task performed. The quality of the query results is an important aspect of preference queries such as top-*k* and skyline, as these query operators are used in many decision making and decision support systems, and are expected to give the most accurate results in terms of helping the user to select the most appropriate decision. Due to the large number of tasks issued by the user and with the time and quality constraints, most studies of processing preference queries for CSDBs have assigned the tasks to either a single or multiple qualified workers. The selection of the most suitable workers to accomplish a task is also an challenging task for the requester, and many crowd-sourced systems achieve this through the use of voting strategies by multiple workers as a tool for aggregating answers from workers. This helps users to make the right decision in terms of choosing qualified workers in order to guarantee the quality of the result. In other cases, workers undergo qualification tests so that low-quality workers will not be assigned the tasks. Some systems rely on the worker's skills and experience in the field of the query (Li *et al.*, 2016). Here, the choice of the most appropriate worker has a high impact on the quality of the results produced by the crowd platform. More recently, several studies have addressed the issue of processing skyline queries on a database with missing data, and investigated the impact of these missing data (Gulzar *et al.*, 2019b; Gulzar *et al.*, 2019c). This problem adds another challenge to the processing of queries in crowd-sourced enabled databases. Some studies have suggested that workers can contribute to estimating the missing values in the crowd, and hence provide a complete answer to the user. This is accomplished by exploiting the implicit relationship between the databases of the crowd. Some notable works that address the issue of preference queries on crowd-sourcing incomplete databases are those by Lofi *et al.* (2013), Lee *et al.* (2016), Swidan *et al.* (2018), Miao *et al.* (2019), and Swidan *et al.* (2020). We suggest that significant research effort should be directed towards addressing the challenges of incomplete and uncertain data when processing queries in a CSDB. Issues such as monetary cost, time latency, and accuracy of results for uncertain and incomplete data should be considered. In the modern era, big data has become a rich area of research, and there are many research opportunities associated with the issue of query processing that could be explored. In this context, data are extremely diverse and the quantities are growing rapidly. Working on big data needs powerful machines and vast human resources to carry out tasks, which makes the process very expensive and takes a relatively long time. Although some studies have concentrated on big data processing, studies have been limited to certain pre-processing tasks such as data cleaning (Wang *et al.*, 2014), and data labelling (Mozafari *et al.*, 2014). Hence, a great deal of work is needed to resolve the problems with data processing in the context of big data. When working with big data with incomplete and uncertain values, extensive computing resources and human effort are needed to carry out the task of missing value estimation. The use of sampling techniques to estimate the missing values is one of the most effective solutions for providing an accurate estimation with reasonable cost.

Another issue that has been addressed in many studies is the privacy of both the workers and the requester. The requester may not wish to reveal all the details of a given task, and some data are highly confidential. The crowd is open to everyone, and some workers might attempt to access the requester database with other public databases, which could lead to data leakage. Concealing the identity of the requester may negatively affect the quality of the task, as this means that workers are unable to access accurate data. Privacy issues add further challenges to the issue of data processing in CSDBs. Designing an approach that balances the accuracy of the result with the privacy of the requester is an interesting idea that should be investigated. The issue of privacy should also be considered from the worker's perspective; in certain cases, workers may have privacy constraints and may be unwilling to share details such as personal information, location, profession and hobbies, which may be requested by the requester to help them in assigning the task. A procedure is therefore needed to ensure the privacy of the worker during assignment of the task.

Another interesting research area is the processing of skyline queries with dynamic and incomplete data. The presence of incomplete data in most contemporary database applications is inevitable (Gulzar *et al.*, 2018; Babanejad *et al.*, 2020). Furthermore, databases are dynamic in nature, as their states change over time to reflect current information about the applications. Frequent changes to the initial database through insert, delete and update operations and the presence of incomplete data means that skylines derived before these changes took place are invalid for the new state of the database. Thus, generating precise predictions for the missing values of skylines after changes have been made to the database is a challenging process. Blindly estimating the missing values of the entire skylines to identify a new set of skylines is unwise, as not all data items will be affected by the changes made to the database. Thus, an efficient solution for imputing the missing values of skylines after changes have been made in an initial incomplete and dynamic database is urgently needed. This solution should be capable of estimating only those skylines that are affected by the changes to the initial database and which have a high probability of being reported in the skyline results for the new state of the database (Babanejad *et al.*, 2020). We believe that the adoption of machine learning-based techniques such as neural networks and generative adversarial networks for value estimation would be an interesting avenue for research.

Last but not least, it has been shown that it is very difficult to control the quality of the work done on crowd-sourced mobile platforms (Li *et al.*, 2016). This is because crowd-sourcing tasks are influenced by several factors related to mobile platforms, such as long distances between mobile phone users and servers, and limitations on power and storage capacity (Elfaki *et al.*, 2019). In addition, some crowd-sourced tasks require access to spatial information about certain tuples, for instance when collecting information about nearby hotels and labelling them. Hence, a worker may only be able to accomplish such tasks for nearby hotels, and those further away would not be included in the task. It would therefore be very interesting to develop an efficient worker selection model that addresses these issues and to investigate the impact of spatial information on the performance of the task. It would also be valuable to investigate the issue of the worker model in relation to the server assignment model, which might help to overcome the limitations of the worker selection model (Li *et al.*, 2016). A server selection model is required that can assign tasks based on the shortest distance, and can select the nearest available qualified worker to accomplish the task according to the user's objectives.

We believe that the present study is valuable, and will open the way for numerous research opportunities in relation to the issue of processing skyline queries with incomplete data in areas such as crowd-sourcing, big data, dynamic and uncertain databases. We have presented and analysed several well-known studies that have focused on the processing of preference queries, such as skyline and top-*k*, for both complete and incomplete CSDBs. We have also presented and discussed some practical solutions that could be employed to handle missing data in various areas such as big data, uncertain, incomplete and dynamic databases. Finally, this study can be used as a departure point for the many researchers who are interested in exploring the challenges of preference queries in the abovementioned areas.

## 5. Conclusion

In this paper, we have discussed query processing in CSDBs, and have highlighted the factors that affect query processing for these databases. Different types of query processing, including traditional and preference queries, have been described and explained. We have also examined some leading studies related to query processing in the area of CSDBs. We have summarised the existing techniques discussed throughout the paper and presented a detailed discussion of current and future research challenges related to preference queries in CSDBs that could be explored by interested researchers.

## Bios

**Marwa B. Swidan**

*International Islamic University Malaysia, Kuala Lumpur, Malaysia, 0060126513398, marwa_happy@yahoo.com*

Mr Swidan is a PhD student at International Islamic University Malaysia, IIUM, Malaysia. She earned her B.Sc. and M.Sc. degrees in Computer Science from Tripoli University, Libya in 2004 and 2009 respectively. Her research interests include crowdsourcing, data mining/exploration, database systems and skyline queries.

**Ali A. Alwan**

*International Islamic University Malaysia, Kuala Lumpur, Malaysia, 0060173546110, aliamer@iium.edu.my*

Dr Alwan is currently an associate professor at Kulliyyah (Faculty) of Information and Communication Technology, International Islamic University Malaysia (IIUM), Malaysia. He received his Master of Computer Science (2009) and Ph.D. in Computer Science (2013) from Universiti Putra Malaysia (UPM), Malaysia. His research interests include preference queries, skyline queries, probabilistic and uncertain databases, query processing and optimization and management of incomplete data, data integration, location-based social networks (LBSN), recommendation systems, and data management in cloud computing.

**Yonis Gulzar**

*College of Business Administration, King Faisal University, Al Ahsa, Saudi Arabia, 00966 54 571 9118, ygulzar@kfu.edu.sa*

Dr Gulzar is currently an assistant professor at King Faisal University (KFU), Saudi Arabia. He obtained his Ph.D. in Computer Science in 2018 from International Islamic University Malaysia. He received his Master in Computer Science in 2013 from Bangalore University, India. His research interests include preference queries, skyline queries, probabilistic and uncertain databases, query processing and optimization and management of incomplete data, data integration, location-based social networks (LBSN), recommendation systems, and data management in cloud computing.

**Abedallah Zaid Abualkishik**

*American University in The Emirates, Dubai, United Arab Emirates, +971501504130, azasoft1@gmail.com*

Dr Abualkishik received a master's and a Ph.D. degree in Software Engineering. His research interests include software functional size measurement, software functional measures conversion, cost estimation, empirical software engineering, database, Big Data, and Data Science. Currently, he is working as an assistant professor at the college of computer information technology, American University in the Emirates, Dubai, UAE. Also, he is working as a consultant for a regional software development company to prepare an accurate estimation for project deliverables.